\documentclass[10pt, a4paper]{article}

\usepackage{amsmath,amssymb,graphicx,a4wide}
\usepackage{epsfig}
\usepackage[sort&compress]{natbib}
\usepackage{authblk}

\newcommand{\argmax}{\operatornamewithlimits{argmax}}
\renewcommand{\vec}[1]{\mathbf{#1}}

\newcommand{\asmall}{\underline{a}}
\newcommand{\alarge}{\overline{a}}

\renewcommand{\Pr}{P}
\newcommand{\Prob}[1]{\Pr\left(#1\right)}
\newcommand{\CondProb}[2]{\Pr\left( #1 \mid #2 \right)}
\newcommand{\integral}[2]{\int \! #1 \mathrm{d}#2}
\newcommand{\gmap}{G_{\mathrm{MAP}}}

\newcommand{\CondExpect}[2]{\mathbb{E}\left[ #1 \middle\vert #2 \right]}
\newcommand{\Gam}[1]{\Gamma\left( #1 \right)}

\title{Bayesian Inference of Whole-Brain Networks}
\author[1,2,*]{M. Hinne}
\author[2]{T. Heskes}
\author[1]{M.A.J. van Gerven}

\affil[1]{Donders Institute for Brain, Cognition and Behaviour: Centre for Cognition. Radboud University, Nijmegen, The Netherlands}
\affil[2]{Institute for Computing and Information Sciences. Radboud University, Nijmegen, The Netherlands}

\affil[*]{mhinne@cs.ru.nl}
\begin{document}
\maketitle

\begin{abstract}
In structural brain networks the connections of interest consist of white-matter fibre bundles between spatially segregated brain regions. The presence, location and orientation of these white matter tracts can be derived using diffusion MRI in combination with probabilistic tractography. Unfortunately, as of yet no approaches have been suggested that provide an undisputed way of inferring brain networks from tractography. In this paper, we provide a computational framework which we refer to as \emph{Bayesian connectomics}. Rather than applying an arbitrary threshold to obtain a single network, we consider the posterior distribution of networks that are supported by the data, combined with an exponential random graph (ERGM) prior that captures a priori knowledge concerning the graph-theoretical properties of whole-brain networks. We show that, on simulated probabilistic tractography data, our approach is able to reconstruct whole-brain networks. In addition, our approach directly supports multi-model data fusion and group-level network inference.
\end{abstract}


\section{Introduction}

Human behaviour ultimately arises through the interactions between multiple brain regions that together form a network that can be characterized in terms of structural, functional and effective connectivity~\citep{penny:2006uw}. The science of characterizing neural connectivity, that is, to elucidate the wiring diagram of the human brain, has come to be known as {\em connectomics} and is seen as one of the great challenges in neuroscience~\citep{Sporns2005}.

Structural connectivity  presupposes the presence of white-matter tracts that connect spatially segregated brain regions which constrain the functional and effective connectivity between these regions. Hence, structural connectivity provides the scaffolding that is required to shape neuronal dynamics. Furthermore, structural connectivity has been related to several diseases, such as Alzheimer's disease, attention deficit hyperactivity disorder and multiple sclerosis~\citep{He2008,Konrad2010,He2009}, and is therefore also of major importance in clinical neuroscience~\citep{Catani:2007dx}.

Currently, the only way to estimate structural connectivity of whole-brain networks in vivo is through the use of diffusion imaging; a variant of magnetic resonance imaging which measures the restricted diffusion of water molecules, thereby providing an indirect measure of the presence and orientation of white-matter tracts. By following the principal diffusion direction in individual voxels, streamlines can be drawn that represent the structure of fibre bundles connecting separate regions of grey matter. This process is known as deterministic tractography~\citep{Zhang2005,Chung2010}. There are however several problems associated with this deterministic approach as it cannot easily deal with various ambiguous but common fiber configurations such as kissing, crossing and splaying fibre bundles~\citep{Jbabdi:2011fn}.

As a solution to this problem, probabilistic tractography has been introduced~\citep{Behrens2003,Behrens2007,Friman2006,Jbabdi2007}. By repeating the streamlining process multiple times, probabilistic tractography ultimately produces for each voxel a measure of uncertainty about in which other voxel a hypothesized connection will terminate. A benefit of the probabilistic approach is that it can deal with ambiguous fiber configurations by taking streamlining uncertainty into account and produces probability estimates concerning the presence or absence of particular white-matter tracts.

Often, one is not interested so much in particular tracts but rather in whole-brain structural connectivity. Several approaches have been suggested to derive whole-brain networks from probabilistic tractography results~\citep{Skudlarski2008,Iturria-Medina2008,Hagmann2007,Chung2011}. The resulting networks are all reported to exhibit certain graph-theoretical properties, such as a characteristic short path length, clustering, the existence of hubs and modular structure~\citep{Bassett2011,Bullmore2009,Bassett2006,Gong2009,Sporns2006}. Unfortunately, the inference of whole-brain networks from probabilistic tractography estimates remains somewhat ad hoc. Networks are often constructed by counting the number of streamlines that connect regions of interest and interpreting this as an undirected, weighted graph (e.g.~\citep{Zalesky2010}). Alternatively, a binary graph is obtained by simply adding an edge for each pair of regions that have one or more streamlines between them (e.g.~\citep{Hagmann2007,Vaessen2010,Bassett2011}). It is hard, however, to maintain a direct correspondence between connection strength and the probability distribution over tracts that is produced by probabilistic tractography~\citep{Jbabdi:2011fn}.

Another important observation is that the mentioned approaches do not easily support the integration of probabilistic streamlining data with other datasets. This is a particularly relevant topic in contemporary neuroscience since it is assumed by multi-modal data fusion, where data acquired via different recording modalities is integrated to provide a coherent picture of brain function~\citep{Horwitz:2002wv}, and required by group-level inference, where the interest is in estimating a network that characterizes a particular population, e.g. when comparing patients with controls in a clinical setting~\citep{Simpson:2011vg}.

In the following, we provide for the first time a computational framework for the estimation of whole-brain networks from probabilistic tractography outcomes. In our approach, which we refer to as {\em Bayesian connectomics}, we consider the posterior distribution of graphs that are supported by our data, instead of a generating a single network based on an arbitrary threshold. Our approach relies on defining a generative model for whole-brain networks which is inspired by recent work on network inference in systems biology~\citep{Mukherjee2008} and consists of two ingredients. First,  a likelihood model based on a multivariate distribution which views the streamline distributions produced by probabilistic tractography as noisy data. Second, an exponential random graph prior which models prior knowledge concerning the graph-theoretical properties of whole-brain networks. The extension to multi-modal data fusion or group-level inference follows by defining the likelihood model as a product of multivariate distributions. In order to validate our methodology we make use of simulated data which allows us to compare the estimates produced by Bayesian connectomics with ground truth. We show that our approach is able to reconstruct whole-brain networks from simulated data as produced by probabilistic tractography.

\section{Bayesian inference of whole-brain networks}

In this section we derive our Bayesian approach to the inference of whole-brain networks, referred to as Bayesian connectomics. We start by defining the likelihood term for our generative model. Subsequently, an exponential random graph prior will be defined which can be used to capture the essential graph-theoretical properties of whole-brain networks. Finally, we derive a Markov chain Monte-Carlo algorithm to sample from the posterior distribution of whole-brain networks.

\subsection{Likelihood model}

Assume we want to infer the connectivity between $K$ brain regions. That is, we want to find a graph $G \in  \mathcal{G}$ where $\mathcal{G}$ is the family of undirected graphs. The graph $G = (V,E)$ consists of vertices $V = \{V_1,\ldots,V_K\}$ and edges $E \subseteq V \times V$ where loops $(V_i,V_i)$ for $i \in \{1,\ldots,K\}$ are excluded from the edge set. We use $e_{ij}=1$ when $(V_i,V_j) \in E$ and $e_{ij}=0$ when $(V_i,V_j) \notin E$.

We start by considering one region and the possible targets in which a postulated tract may terminate. Probabilistic tractography will produce a distribution over target regions $\vec{n}_i=(n_{i1}, \ldots, n_{iK})^T$ by drawing $S$ streamlines, $N_i = \sum_{k=1}^K n_{ik} \leq S$ of them ending up in a target region. Let $n_{ii}=0$ and assume that all streamlines are drawn independently. The probability of finding a particular distribution $\vec{n}_i$ is expressed as a multinomial distribution
\begin{equation}
    \CondProb{\vec{n}_i}{\vec{x}_i} \propto \prod_{j=1}^K x_{ij}^{n_{ij}} \enspace,
\end{equation} in which $x_{ij}$ with $\sum_j x_{ij} = 1$ represents the probability of drawing a streamline from region $i$ to region $j$. This streamlining probability itself depends on whether or not there actually exists a tract between region $i$ and region $j$. We model the distribution of streamlining probabilities using a Dirichlet distribution
\begin{equation}
    \CondProb{\vec{x}_i}{\vec{a}_i}
    \propto
        \prod_{j=1}^K x_{ij}^{a_{ij}-1}
\end{equation}
where the
\begin{equation}
a_{ij} = e_{ij}\alarge + (1-e_{ij})\asmall
\label{eq:diriparams}
\end{equation}
can be interpreted as pseudo-counts that determine the probability of streamlining from region $i$ to region $j$ when an edge $e_{ij}$ in the underlying graph $G$ is either present $(\alarge)$ or absent ($\asmall$). Because the Dirichlet prior with parameters $\vec{a}_i = (a_{i1},\ldots,a_{iK})$ is conjugate to the multinomial distribution, the posterior distribution of tract probabilities is again Dirichlet such that
\begin{equation}
    \CondProb{\vec{x}_i}{\vec{n}_i, \vec{a}_i} \propto P(\vec{n}_i \mid \vec{x}_i) P(\vec{x}_i \mid \vec{a}_i) \propto \prod_{j=1}^K x_{ij}^{n_{ij} + a_{ij} - 1}\enspace.
\end{equation}

Let $\vec{N}=\left(\vec{n}_1; \ldots; \vec{n}_K\right)$ represent the probabilistic tractography data, $\vec{X}=\left(\vec{x}_1; \ldots; \vec{x}_K\right)$ the streamlining probabilities, and $\vec{A}=\left(\vec{a}_1; \ldots; \vec{a}_K\right)$ the parameters of the Dirichlet distribution, for all brain regions $1 \leq i \leq K$ combined. The likelihood of the graph $G$ is then expressed as
\begin{eqnarray}\label{eq:multinom}
    \CondProb{\vec{N}}{G}
        &=& \integral{
                \CondProb{\vec{N}}{\vec{X}}\CondProb{\vec{X}}{\vec{A}}}{\vec{X}}\notag\\
                &=& \prod_i \left[ \frac{N_i!}{\prod_{j} n_{ij}!}
            {
               \Gamma\left(\sum_j a_{ij}\right)
                \over
                \Gamma\left( \sum_j \left( a_{ij} + n_{ij} \right) \right)
            }
           \prod_j
            {
                 \Gamma\left( a_{ij} + n_{ij} \right)
                 \over
                \Gamma\left( a_{ij} \right)
            } \right]\enspace,
\end{eqnarray}
where we have replaced $\vec{A}$ with $G$ on the left-hand side to emphasize that we are interested in a probability distribution over a set of graphs. Note that $\vec{A}$ and $G$ are easily exchanged through Eq.~\eqref{eq:diriparams}. The second step in Eq.~(\ref{eq:multinom}) follows by recognizing $\CondProb{\vec{N}}{G}$ as a product of compound Dirichlet-Multinomial distributions, also known as multivariate P\'olya distributions~\citep{Minka:2003ve}. This leads to the log-likelihood
\begin{equation}\label{eq:loglikelihood}
   L \equiv  \log \CondProb{\vec{N}}{G} = \sum_i \left[ \log \frac{N_i!}{\prod_{j} n_{ij}!}   +
             \log \frac{\Gam{\sum_j a_{ij}}}{\Gam{\sum_j (a_{ij}+n_{ij})}}+
            \sum_{j} \log \frac{\Gam{a_{ij}+n_{ij}}}{\Gam{a_{ij}}}\right]
\enspace.
\end{equation}

\subsection{Exponential random graph prior}

We would like to define a prior such that, when sampling from the prior, we generate reasonable whole-brain networks. Here, reasonable is taken to mean that the networks should respect graph-theoretical properties that have been inferred by previous studies in connectomics. One way to construct a prior that has this property is using the theory of exponential random graph models (ERGMs)~\citep{Holland:1981ww,Strauss:1986va,Wasserman:1996th}. ERGMs, developed in the context of social network analysis~\citep{Snijders2002,Snijders2006} and systems biology~\citep{Mukherjee2008}, have recently been used to fit models of whole-brain networks~\citep{Simpson:2011va,Simpson:2011vg}. However, they have not been used before as priors in a generative model of whole-brain networks.

The exponential random graph prior is defined as a log-linear distribution
\begin{equation}
    \Prob{G} \propto \exp \Bigl(\lambda\sum_i w_i f_i(G)\Bigr)\enspace,
\end{equation} in which $\lambda$ is a strength parameter and the parameters $w_i$ tune the relative strengths of the individual concordance functions $f_i:\mathcal{G}\rightarrow \mathbb{R}$ that capture several graph-theoretical properties of whole-brain networks. 

For illustrative purposes, we will use the prior in order to capture the small-world phenomenon that occurs so frequently in brain networks~\citep{Sporns:2010tk}. That is, we wish to generate networks that have short average path length and a large clustering coefficient. This can be achieved by starting with a graph whose neighboring vertices are connected and rewiring edges at random with a certain probability~\citep{Watts:1998db}. The choice of functions $f_i$ is restricted however by the Markov chain Monte Carlo sampling algorithm since, as we will note in Section~\ref{MCMC}, the change after each sampling step must be calculated efficiently. Unfortunately, this prevents us from using average path length as as a concordance function since the change in path length due to an edge mutation cannot be calculated locally. It requires an all-pairs shortest paths algorithm, for which the fastest -- Dijkstra's algorithm -- is known to have time complexity $O(|E|\cdot|V| + |V|^2\log |V|)$ on sparse graphs~\citep{Dijkstra1959}. Instead, as a surrogate model, we use a prior that prefers graphs which are characterized by a small number of short edges. Let $d_{ij}$ be the Euclidian distance between vertices $i$ and $j$, then the prior is defined as
\begin{equation}
    \Prob{G}\propto \exp\left( \beta f(G)  \right)
    \label{prior}
\end{equation}
with $\beta = \lambda \cdot w$ and concordance function $f(G) = -\sum_{i,j} e_{ij} d_{ij}$ which assumes that the edge length follows an exponential distribution. In Section~\ref{simulation}, we will show that this prior prefers graphs which exhibit the small-world property.

\subsection{Sampling from the posterior distribution}
\label{MCMC}

By using the likelihood and the prior in conjunction with Bayes' rule, we obtain the following posterior distribution over whole-brain networks:
\begin{equation}
    \CondProb{G}{\vec{N}}\propto \CondProb{\vec{N}}{G}\Prob{G}\enspace.
\end{equation}
Since the posterior $\CondProb{G}{\vec{N}}$ cannot be calculated analytically, we use Metropolis Markov chain Monte Carlo (MCMC) sampling to approximate this distribution~\citep{Mukherjee2008}. The acceptance of each sample $G'$ in the sampling chain is determined by the ratio
\begin{equation}\label{eq:acceptance}
    \alpha=\frac{\CondProb{G'}{\vec{N}}}{\CondProb{G}{\vec{N}}}\enspace.
\end{equation} A proposed graph becomes a new sample with probability $\min(1,\alpha)$. The acceptance probability of a suggested sample can be calculated as
\begin{equation}\label{eq:deltaposterior}
    \log\alpha = \Delta L_{kl} + \lambda\sum_i w_i \Delta f_i(G)\enspace,
\end{equation} with $\Delta L_{kl}$ the change in log-likelihood after flipping edge $e_{kl}$ and $\Delta f_i(G)=f_i(G')-f_i(G)$.
This requires that we can efficiently update both the likelihood and the prior for new samples in the Markov chain.

The change in log-likelihood as a consequence of the flipping of an edge $e_{kl}$ to $1 - e_{kl}$ is defined as 
\begin{equation}\label{eq:deltaloglikelihood}
    \Delta L_{kl}= \log \CondProb{\vec{N}}{G'} - \log \CondProb{\vec{N}}{G}\enspace,
\end{equation} with the sole difference that the parameter $a_{kl}$ associated with $G'$ and denoted by $\tilde{a}_{kl}$ is given by $\tilde{a}_{kl}= \tilde{a}_{lk} = (1-e_{kl}) \alarge + e_{kl} \asmall $ with $e_{kl}$ the edge as defined in $G$. Plugging~\eqref{eq:loglikelihood} into~\eqref{eq:deltaloglikelihood} yields
\begin{eqnarray}
    \Delta L_{kl}
    &=&
    \log \left[
        {
            \Gam{\tilde{a}_{kl}+n_{kl}}
            \over
            \Gam{a_{kl}+n_{kl}}
        }
    \right]
    +
    \log \left[
        {
            \Gam{\tilde{a}_{lk}+n_{lk}}
            \over
            \Gam{a_{lk}+n_{lk}}
        }
    \right] +
    \log \left[
        {
            \Gam{\sum_j \tilde{a}_{kj}}
            \over
            \Gam{\sum_j a_{kj}}
        }
    \right]+
    \log \left[
        {
            \Gam{\sum_j \tilde{a}_{lj}}
            \over
            \Gam{\sum_j a_{lj}}
        }
    \right]
    \notag\\
    &&
    -
    \log \left[
        {
            \Gam{\sum_j (\tilde{a}_{kj}+n_{kj})}
            \over
            \Gam{\sum_j (a_{kj} + n_{kj})}
        }
    \right]
    -
    \log \left[
        {
            \Gam{\sum_j (\tilde{a}_{lj}+n_{lj})}
            \over
            \Gam{\sum_j (a_{lj} + n_{lj})}
        }
    \right]
        -
    2 \log \left[
        {
            \Gam{\tilde{a}_{lk}}
            \over
            \Gam{a_{lk}}
        }
    \right]
    \enspace.
    \label{deltaL}
\end{eqnarray}

The change in the log prior as a consequence of the flipping of an edge $e_{kl}$ to $1 - e_{kl}$ depends on the employed concordance function(s). The concordance function used in (\ref{prior}) factorizes over individual edges:
\begin{equation}
    P(G) \propto \prod_{i,j} \exp\left[ - \beta e_{ij} d_{ij}\right]
\end{equation}
such that $\Delta f(G) =  (2e_{kl} -1)d_{kl}$.

All configurations of the $2^{N(N-1)/2}$ possible graphs have a probability greater than zero to be constructed thus guaranteeing that the Markov chain is irreducible. The collection of accepted samples $\{G^{(1)}, \ldots, G^{(T)}\}$ forms an approximation of the posterior $\CondProb{G}{\vec{N}}$. When we desire to generate a whole-brain network, the maximum a posterior graph
\begin{align}\label{eq:map}
    \gmap
        &= \argmax_G \Bigl\{\log \CondProb{\vec{N}}{G} + \log \Prob{G}\Bigr\}
\end{align}
can be used. Alternatively, the samples can be used to estimate posterior probabilities of network features, such as the existence of a tract. Assuming the Markov chain has converged, the posterior probability of a single tract is given by
\begin{equation}
    \CondExpect{e_{ij}}{\vec{N}} = \frac{1}{T} \sum^T_{t=1} e_{ij}^{(t)}\enspace.
\end{equation} Other quantities may be estimated in a similar fashion.

\subsection{Dealing with multiple datasets}

As pointed out previously, probabilistic streamlining data need not be collected in isolation. It may be collected in conjunction with other data such as resting-state fMRI or MEG data or it may be collected for multiple subjects. We show that our approach quite naturally supports multi-modal data fusion and/or group-level inference.

Suppose we have acquired $M$ datasets $\vec{Y}_1,\ldots,\vec{Y}_M$.  The goal is to derive the graph $G$ which best explains these data. The posterior for $G$ will be of the form
\begin{equation}
 \CondProb{G}{\vec{Y}_1,\ldots,\vec{Y}_M}\propto \CondProb{\vec{Y}_1,\ldots,\vec{Y}_M}{G}\Prob{G}\,.
\end{equation}
Conditional on $G$, we assume that the datasets are independent, such that we have the following factorized representation of the posterior:
\begin{equation}
 \CondProb{G}{\vec{Y}_1,\ldots,\vec{Y}_M}\propto \left[ \prod_{i=1}^M \CondProb{\vec{Y}_i}{G}\right] \Prob{G}\,.
 \label{factor}
\end{equation}
In case of multi-modal data fusion this requires that we are able to write down a likelihood model for each of the datasets $\vec{Y}_i$. In case of group-level inference, $\vec{Y}_i$ represents probabilistic streamlining data for subject $i$. Note that this is not the same as simply summing the streamlines, by virtue of Eq.~(\ref{eq:multinom}). It follows from the factorization implied by~(\ref{factor}) that, in order to run the MCMC algorithm at the group level, all that is required is to use the following term for the change in log-likelihood when computing the acceptance probability:
\begin{equation}
\Delta L_{kl} = \sum_{i=1}^M \bigl[ \log \CondProb{\vec{N}_i}{G'} - \log \CondProb{\vec{N}_i}{G} \bigr] \,.
\end{equation}

\section{Simulation study}
\label{simulation}

\begin{figure}[ht]
      \centering      \includegraphics[width=0.8\textwidth]{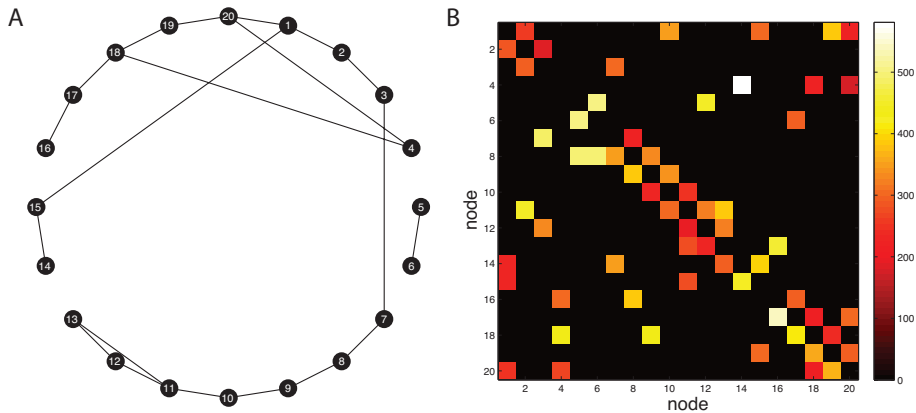}
      \caption{Simulated data. A. Small-world graph $G$. B. Probabilistic streamlining data generated for graph $G$ indicating the number of  streamlines drawn between nodes $i$ and $j$.}
      \label{sim1}
\end{figure}

In our simulation study, we used the Watts \& Strogatz model~\citep{Watts:1998db} in order to generate a small-world graph $G$ consisting of 20 edges and 20 vertices, laid out in a circular arrangement that defines the physical distance between the nodes. For this ground-truth network, we generated `probabilistic streamlining' data as follows. For each node $i$  a streamlining probability vector $\vec{x}$ was constructed by drawing from a Dirichlet distribution with parameters $\vec{a}_i = (a_{i1},\ldots,a_{iN})$. Parameters were set to zero except for those $a_{ij}$ for which $(i,j) \in E$ (the true tracts) as well as for a randomly selected $a_{ik}$ for each $1 \leq i \leq 20$ (representing a false positive due to kissing, crossing or splaying fiber bundles). These parameters were all set equal to 20. Subsequently, for each node $i$ the streamlining probability vectors $\vec{x}_i$ were used to generate 1000 streamlines drawn from a multinomial distribution. Figure~\ref{sim1} shows $G$ together with the generated probabilistic streamlining data $S$.

Next, we wanted to determine whether the ground-truth network could be retrieved using Bayesian connectomics. That is, can we reconstruct $G$ from probabilistic streamlining data $S$ using our MCMC algorithm, does the exploitation of prior information help in retrieving the original network, and does it outperform a conventional thresholding approach? We ran the MCMC algorithm using five parallel chains, a burn-in of 2000 steps and an effective sample size of $8000$. For the hyperparameters, we used $\alarge = 20$ and $\asmall = 15$ to indicate that absent edges may still have a substantial probability of being streamlined. For our prior, we used $\beta = 2$ as this generated networks whose number of  edges agreed with what one expects if the underlying network is generated by the Watts \& Strogatz model.

\begin{figure}[ht]
      \centering      \includegraphics[width=0.55\textwidth]{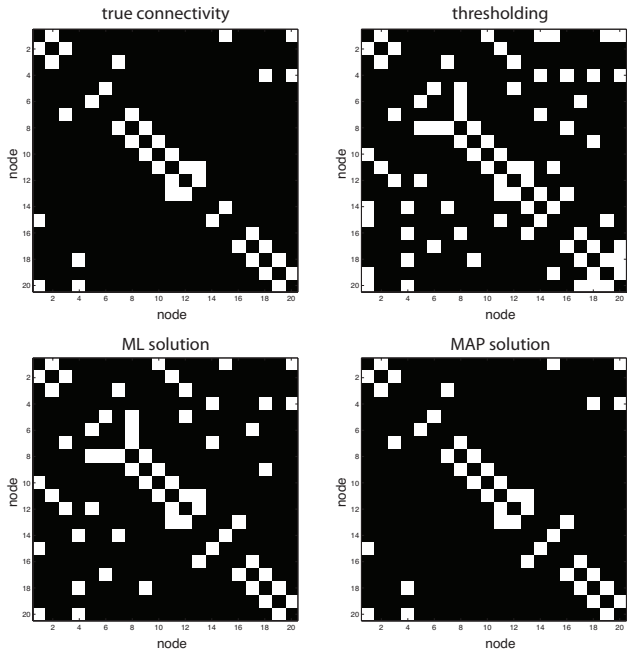}
      \caption{Comparison between the true connectivity and the connectivity obtained with a thresholding approach, a sample from the prior, the maximum likelihood (ML) solution and the maximum a posteriori (MAP) solution afforded by Bayesian connectomics.}
      \label{sim2}
\end{figure}

Figure~\ref{sim2} shows the true connectivity as well as the reconstruction results for a conventional thresholding approach (containing those edges for which the number of streamlines exceeds zero), the maximum likelihood solution, which is obtained by running the MCMC algorithm with a flat prior, and the maximum a posteriori solution, which is obtained by running the MCMC algorithm with an informed prior. The latter is the procedure we refer to as Bayesian connectomics. Clearly, both the thresholding approach and the maximum likelihood solution suffer from the false positives introduced during the streamlining process. The MAP solution, in contrast, is able to retrieve the true connectivity almost perfectly well. Note that this cannot be explained by the use of the prior alone since the prior assumes that the empty graph is the most probable graph and samples from the prior are not likely to recover the long-range connections.

\section{Discussion}

Diffusion imaging promises to reveal the the wiring diagram of the human brain in vivo as achieved by probabilistic tractography. So far, the generation of whole-brain networks from probabilistic tractography data has relied on heuristic thresholding methods. In this paper we propose Bayesian connectomics as a principled framework that may replace these thresholding methods. The framework allows for the exploitation of prior knowledge when deriving the true connectivity based on inherently noisy probabilistic streamlining data. An additional advantage of Bayesian connectomics is that it easily allows dealing with multiple datasets as is the case with multi-modal data fusion or group-level inference. We have shown that our approach outperforms thresholding in retrieving the true connectivity underlying simulated probabilistic streamlining data.

In the simulation, we made a number of assumptions. First, we assumed that errors during probabilistic streamlining arise through the existence of a restricted number of false positives which arise due to the existence of a few kissing, crossing and splaying fibers. It would be interesting to examine how our approach behaves under other, possibly more realistic, assumptions about how probabilistic streamlining data is generated. Furthermore, we assumed that the hyperparameters $\alarge$, $\asmall$ and $\beta$ were known. The hyperparameters of the Dirichlet distribution are unknown in practice, but these could be chosen based on the distribution of the number of streamlines that are generated during the streamlining process. The parameter $\beta$, which determined the influence of the prior was chosen such that samples from the prior corresponded to the small-world networks we were interested in. In order to construct priors which work well in practice, various other concordance functions can be employed, whose parameters can be estimated on existing data using maximum likelihood approaches~\citep{Hunter:2008va,Simpson:2011va}.

In this paper, we have laid the foundations for a Bayesian approach to connectivity analysis. What remains is to validate the approach not only on simulated data but also on empirical data. This not only requires the reconstruction of whole-brain networks from diffusion imaging data, but also an assessment of their validity on independent data. One may think here of determining how well reconstructed networks correspond to networks estimated from resting-state fMRI or from tracer data~\citep{Jbabdi:2011fn}. Eventually, we would like to use our approach to determine how differences between (groups of) people can be understood in terms of variations in their underlying wiring diagrams.

\bibliographystyle{apalike}
\bibliography{library}

\end{document}